\documentclass[12pt]{article}
\usepackage[margin=1in]{geometry}
\usepackage{amsmath,amssymb,amsfonts}
\usepackage{graphicx}
\usepackage{float}
\usepackage{url}
\usepackage{authblk}
\usepackage{algorithm}
\usepackage{algorithmic}

\date{} 

\begin{document}
\title{A Statistical Side-Channel Risk Model for Timing Variability in Lattice-Based Post-Quantum Cryptography}

\author[1]{Aayush Mainali}
\author[2]{Sirjan Ghimire}

\affil[1]{Department of Computer Science and Engineering, Amrita School of Computing, Bengaluru, Amrita Vishwa Vidyapeetham, India}
\affil[2]{Bal Kalyan Vidhya Mandir, Biratnagar, Nepal}

\maketitle
\begin{abstract}
Timing side-channels are an important threat to cryptography that still needs to be addressed in implementations, and the advent of post-quantum cryptography raises this issue because the lattice-based schemes may produce secret-dependent timing variability with the help of complex arithmetic and control flow. Since also real timing measurements are affected by environmental noise (e.g. scheduling effects, contention, heavy tailed delays), in this work a scenario-based statistical risk model is proposed for timing leakage as a problem of distributional distinguishability under controlled execution conditions. We synthesize traces for two secret classes in idle, jitter and loaded scenarios and for multiple leakage models and quantify leakage with Welch's t-test, KS distance, Cliff's delta, mutual information, and distribution overlap to combine in a TLRI like manner to obtain a consistent score for ranking scenarios. Across representative lattice-based KEM families (Kyber, Saber, Frodo), idle conditions generally have the best distinguishability, jitter and loaded conditions erode distinguishability through an increase in variance and increase in overlap; cache-index and branch-style leakage tends to give the highest risk signals, and faster schemes can have a higher peak risk given similar leakage assumptions, allowing reproducible comparisons at an early design stage, prior to platform-specific validation.
\end{abstract}

\textbf{Keywords:} post-quantum cryptography; lattice-based KEM; timing side-channel; leakage modeling; statistical distinguishability; mutual information; Kolmogorov--Smirnov distance; TLRI risk score

\section{Introduction}
Post-quantum cryptography (PQC) is being adopted to insure against long-term secret information from adversaries with quantum capabilities. While many PQC candidates are highly secure mathematically, to make PQC a practical choice, they also need to resist to deployments at an implementation level. Timing side-channels are of particular importance because they may leak information due to small differences in execution time caused by secret-dependent control flow, the (variable) latency of arithmetics or memory accesses. In the case of lattice-based constructions, performance-critical operations such as polynomial arithmetic, transforms, encoding/decoding and validation logic can inadvertently introduce measurable timing variability even with the assumption that the underlying protocol is formally secure.

One of the main challenges in the timing of side-channel analysis is that the actual timing measurements are influenced not only by effects of the secret, but also by the execution environment. Scheduling jitter, interrupts, cache contention and other system-level behaviors result in changes to the timing distributions, which will affect the leakage, masking it or, in some cases, multiplying it. As a result, the practical question is often not whether leakage exists in principle, but rather how consistently an attacker can look for secrecy differences for realistically-ranging variability and limited samples. This leads to the need for methods for assessing-malexia in which environment noise is explicitly added, and which allow comparison of the leakage risk for different algorithmic styles, platforms, and assumptions about the threat.

Existing side channel research often shows convincing examples of effective attacks or suggests specific counter-measures, however, much fewer methods feature a unified and reproducible methodology to quantify and rank risk for timing leakage in multiple environments and leakage mechanisms. In particular, the variation in timing is frequently treated as a nuisance variable rather than as a first-class modelling aim, and assessments can be scattered over scheme households and measuring measurement environments. This restricts the ability of practitioners to reason in a uniform manner about which combinations of scheme, environment and leakage mechanism produce the greatest practical risk.

To address these gaps, a scenario-based statistical risk model for the timing of leakage is introduced in this paper for lattice-based PQC. We define a scenario to consist of the combination of a scheme-level baseline timing scale, an environment regime (idle, jitter, loaded), a leakage model that describes the modulation of timing secrets and the leakage strength parameter which controls the magnitude of the effect. For each scenario, we obtain timing traces for two balanced secret classes and measure distinguishability with a set of complementary statistical and information-theoretic metrics. Finally, we combine these metrics into a TLRI-type composite score which assists in consistent ranking of the risk of leakage for scenarios.

The main contributions of this work are:
\begin{itemize}
    \item a re-playable scenario-based simulator that separates the baseline cost, environment noise, and secret dependent leakage;
    \item a multi-metric evaluation pipeline based on mean-based, distributional and dependency measures;
    \item a bounded composite risk score to allow cross scenario comparison; and
    \item empirical comparative study of representatives of lattice based KEM families including Kyber, Saber and Frodo and various leakage mechanisms
\end{itemize}
Together, these contributions offer an early stage risk modelling tool that would help to guide where deeper platform specific measurements and countermeasures efforts should be prioritized.

\section{Literature Review}

\subsection{Timing Side-Channels: Foundations and Modern Threat Surfaces}
Timing side channels are some of the oldest and most significant classes of implementation vulnerabilities in cryptography, as they illustrate how even with secure underlying mathematics, differences in the time to perform some computer operation may reveal something sensitive. Classical work showed that timing differences arise due to the use of conditional branches, variable-latency arithmetics, and memory access patterns can be used to recover secrets within popular cryptosystems and the foundational adversarial principle, ``time is an information channel'' \cite{b1}. This is a basic understanding that is still very pertinent in post-quantum cryptography (PQC) where implementations frequently involve the use of large polynomial arithmetics, structured transforms and complex decoding paths which can inadvertently create secret dependent variability.

In parallel, modern systems enlarge the timing attack surface to include more than user-space cryptographic routines. Practical attacks have also demonstrated that timing can be used to disclose characteristics of the privilege address space and layout randomization properties, to illustrate the impact of the operating system and microarchitectural behavior on the timing distributions that can be measured \cite{b2}. In PQC deployments, these system-level effects are important because timing measurements can be impacted by scheduling effects, interrupts, cache contention, and power management behavior that can both mask as well as amplify leakage signals based on the environment.

A lot of work is also done for memory timing and cache timing leakage by treating them as first-class threats through proposing different mitigation mechanisms to change the execution with lower timing dependence of memory access behavior. Techniques such as mitigating memory timing side-channels through changes to the execution structure \cite{b3} and cache locking with fine-grained partitioning \cite{b4} reflect the fact that mitigating timing side-channel attacks is often required at the memory hierarchy level, rather than at the algorithmic level. These insights inspire evaluation methods that explicitly account for variation of platforms and environments when reasoning about timing leakage risk.

\subsection{Side-Channel and Fault Attacks in the Post-Quantum Era}
As PQC candidates were progressing towards standardization, research increasingly showed that post quantum primitives are vulnerable to both passive and active implementation attacks. Fault-injection analyses against Round 3 KEM candidates find that well-designed induced faults can result in only partial or complete secret recovery as well as counter security presumptions when the implementations do not come with proper countermeasures and verification measures \cite{b5}. This work focusing on timing leakage is complementary to timing leakage studies, and provides an important reminder that the security a system offers in practice is highly dependent upon the complete stack of implementation, and also the capabilities of the attacker to control execution conditions.

Beyond fault models, PQC has been proven to be susceptible to power and electromagnetic (EM) analysis; including generic attack frameworks that are applicable to several KEM constructions. One of the messages significant is that some re-encryption or decapsulation behaviors can result in consistent side-channel features exploitable in power/EM analysis even on what at first glance seems like an industrial-strength protocol design \cite{b6}. Code based schemes have also been explored within frameworks of side-channel key recovery techniques to formalise the way that leakage is exploited to recover secret structure \cite{b7} which reinforces the notion that analysis of leakage should not be restricted to a single family of PQC.

Concrete demonstrations further emphasize the fact that leakage may occur in diverse PQC families. For example, a power side-channel vulnerability has been demonstrated on HQC variants, which indicate that the decoding and other operations can reveal secret dependent signals under realistic measurement assumptions \cite{b8}. What is more, plaintext-checking side-channel attacks point out that decapsulation logic and validity checks can leak information on greater complexity than just "bit by bit" results allowing for more powerful forms of recovery \cite{b9}. Collectively, these different works suggest that PQC security assessments should involve passive leakage and active perturbation and multiple leakage points should be considered over encoding, transform, and validation phases.

\subsection{Leakage in Lattice-Based Constructions: Encoding, NTT, and Algorithmic Structure}
Lattice based schemes are most common for standardized PQC, but implementations of Lattice schemes are prone to several leakage hot spots. A major theme is that message encoding and decoding steps may leak through single-trace side channel observations even in the case of large and noisy overall computation. Single-trace attacks on message encoding in lattice-based KEMs: Short-lived but structured operations can disclose sensitive information with surprisingly low measurement requirements \cite{b10}. Similar understandings can be made for transform-based arithmetic as more practical single-trace attacks for the number theoretic transform (NTT) demonstrate that transform structure, table lookups and memory behaviour become exploitable features, especially when operations are not careful constant-time and constant-memory \cite{b11}.

Scheme design choices may have an impact on the security and the attack surface width. FrodoKEM for example is based on LWE without the use of structured ring arithmetic and makes a tradeoff between performance and conservative assumptions, but still relies on the system-level resilience of the implementation and platform behavior \cite{b12}. Notably, end-to-end key recovery on FrodoKEM using Rowhammer shows that even if the timing leakage can be controlled to ensure that both the microarchitectural fault mechanisms can reveal secrets if there is no integrity protection over memory integrity \cite{b13}. This reiterates the fact that PQC deployments need holistic defenses that cover timing, power, cache and fault dimensions.

Other candidates based on lattice focus more on reducing attack surface by way of simplicity in design and implementation. NTRU Prime specifically aims at diminishing structural weaknesses and opportunities for practical attacks over previous NTRU-like constructions \cite{b14}. Meanwhile, work discussing the use of lattice codes in CRYSTALS-Kyber helps to explain a link between design choices and efficiency and correctness and also why implementation details such as sampling, compression and polynomial arithmetic are central to a working security \cite{b15}. In addition, robustness goals beyond standard confidentiality are a concern in PQ settings; robust and anonymous encryption constructions highlight the fact that real-world deployments may require privacy properties that interact with the implementation behavior of such schemes \cite{b16}, and analysis of anonymity properties across Round 3 KEMs shows that even metadata leakage can be relevant to practice \cite{b17}.

\subsection{Hardware and Accelerator Implementations: Performance--Security Trade-offs}
Given the computational intensity of PQC, hardware-acceleration is a popular deployment direction, yet hardware brings its own leakage and design trade-offs with it. Configurable crypto-processors for lattice based protocols illustrate the impact of parameterization and architecture trade-offs in intra-processor analysis to improve throughput and flexibility and radically question one's confidence in side-channel analyses against shared functional units and memory interfaces \cite{b18}. Similarly, the high-level synthesis (HLS) based implementations focus on fast design space exploration and portability at the cost of careful considerations of constant-time properties, memory access patterns, in order to avoid synthesizing the unintended leakage \cite{b19}.

Dedicated coprocessors that are integrated with general purpose CPUs are another important approach. A post-quantum cryptography coprocessor for a RISC-V core, this example demonstrates how operations in PQ can be offloaded and the programmable context of the system is preserved and is very attractive for embedded and edge deployments \cite{b20}. FPGA studies on efficiency and security in key establishment mechanisms of quantum resistance clearly point to the role of throughput, latency, and resource utilization constraints in leading to design architectural shortcuts that might impact leakage behavior, unless they are controlled explicitly \cite{b21}. These works collectively are motivated to have evaluation frameworks to compare risk of leakage across multiple environments and hardware design points as opposed to assuming stable execution profile.

Constant time and constant latency hardware arithmetic is often put forward as a fundamental mitigating tactic. Area-optimized constant time hardware implementations to polynomial multiplication It is an interesting finding that it is possible to address the computation of polynomial multiplication in constant time (even when tightly limited in terms of resources), but it needs to be taken into account that the design will have to deal with both the computation and the memory interface timing \cite{b22}. Taken together, the hardware literature implies that one can expect to see performance improvements by doing so, but that leakage resilience would have to be considered a co-equal design goal, and needs to be verified under real execution conditions, rather than under ideal conditions.

\subsection{From Isolated Attacks to Risk Quantification Under Timing Variability}
A recurring limitation throughout the side-channel literature is that many studies introduce highly effective attacks or targeted countermeasures, while fewer approaches are available to unify a way of quantifying and comparing the leakage threat across environments and noise regimes as well as implementation styles. Foundational timing attack principles explain the mechanism of leakage \cite{b1}, while timing studies at the OS level reveal the introduction of variability in the platforms that result in the alteration of observable distributions \cite{b2}. For PQC, where computations are large and heterogeneous, the practical challenge can often be not whether or not there exists leakage in principle, but rather how reliably an adversary would be able to tell secrets apart given realistic timing variability and the availability of limited samples.

Results of attacking different PQ families show that the distinguishers can take advantage of different statistical signals with respect to the leakage point. Generic power/EM analyses on KEMs prove that there are consistent leakage features that remain across constructions.\cite{b6}, while plaintext-checking attacks have revealed that there are richer leakage channels that are able to encode more information than just binary outcomes.\cite{b9} Single-trace studies also suggest that the attackers do not necessarily need large traces if a sensitive subroutine is structured enough \cite{b10} and work on transform-level leakage also suggests the existence of certain arithmetic patterns to dominate the signal \cite{b11}. Meanwhile, memory and cache mitigation studies focus on the existence of a leakage frequently came from the interaction between the computation and memory hierarchal behavior \cite{b3,b4}.

These observations motivate risk modeling approaches which treat leakage as a distributional phenomenon which is influenced by both environment-independent leakage effects which are secretdependent and environment-dependent noise sources. Hardware and FPGA studies as such demonstrate that the implementations and platform selections have a material impact on the timing, side-channel characteristics \cite{b20,b21} and even conservative lattice designs are still weakened by practical system level threats \cite{b12,b13}. Therefore, a direction that seems to emerge from the literature, is from isolated leakage demonstrations to comparative and scenario-based quantification of leakage, or the ability to rank risk across schemes, environments, and leakage mechanisms, in a consistent way.

\subsection{Identified Gaps and Motivation for This Work}
\begin{itemize}
\item \textbf{Lack of unified, comparable risk score across environments:} Several works show attacks or mitigations, but do not give a unified exercise to compare leakage risk consistently under idle, jittery, and loaded execution conditions \cite{b1,b2}.
\item \textbf{Limited attention to timing variability as a first class modeling target:} System noise sources (interrupts, DVFS, cache contention) cause variations in the observability of leakage, but timing variability can be seen as a 'nuisance' and not explicitly modeled \cite{b2,b4}.
\item \textbf{Fragmented evaluation across PQ families and leakage mechanisms:} Evidence exists across a variety of lattice-based, code-based, and hybrid-based designs-but evaluation typically becomes siloed \cite{b6,b7,b8}, which in turn makes comparison between schemes hard to accomplish.
\item \textbf{Insufficient scenario-driven synthesis of leakage signals:} Single trace leakage, plaintext check leakage and microarchitectural faults provide different attack routes, but little is done to integrate this in a single comparative view of overall implementation risk \cite{b9,b10,b11,b13}.
\item \textbf{Need for reproducible, early-stage assessment tools:} Hardware and accelerator works demonstrate large design spaces with security tradeoffs \cite{b18,b19,b20,b21}, but practitioners benefit from a reproducible evaluation pipeline that models leakage and noise prior to locking in a particular platform.
\end{itemize}

\section{Methodology}

\subsection{Overview of the Experimental Design}
\subsubsection{Scenario definition and experimental objective}
The simulation aims at measuring the timings side-channel distinguishability between two secret classes and controlled, repeatable conditions. A scenario is a single experimental unit and it is defined as
\[
\mathcal{S}=(\text{scheme},\ \text{environment},\ \text{leak model},\ \alpha),
\]
In which scheme indicates an artificial cost model of the base, environment indicates the regime of external noise and contention, leak model indicates the leak data of the form of secret-dependent timing modulation, and $\alpha$ is a scalar leak strength. In both cases, we create $N$ timing traces, which are independent, and evaluate whether the induced timing distribution differs between secret classes $s\in\{0,1\}$. The result is a system of complementary statistical and information theoretic measures (Welch $t$, KS-$D$, Cliff's $\delta$, mutual information, overlap) and a TLRI-style composite score, to allow comparable ranking of scenarios. Significantly, the simulator is applied in comparative risk modeling: the effects of environment and leakage assumptions on detectability are investigated, as opposed to making platform-specific measurements.

\begin{table}[h]
\centering
\small
\begin{tabular}{ll}
\hline
\textbf{Component} & \textbf{Role in a scenario $\mathcal{S}$} \\
\hline
scheme & baseline cycle scale $B$ and noise/structure parameters \\
environment & external regime: idle / jitter / loaded \\
leak model & secret-dependent timing modulation $\Delta(\cdot)$ \\
$\alpha$ & scalar multiplier controlling leakage strength \\
\hline
\end{tabular}
\caption{Scenario definition used throughout the methodology.}
\end{table}

\begin{figure}[H]
    \centering
    \includegraphics[width=1\linewidth]{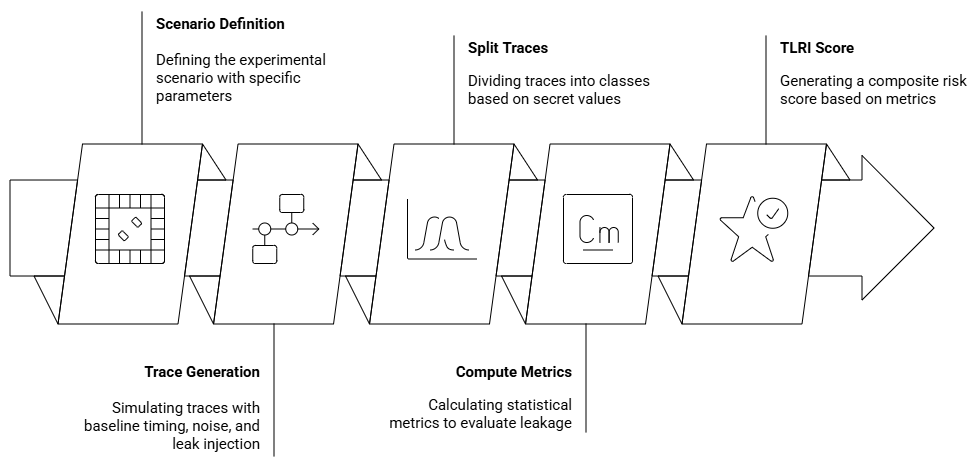}
    \caption{Scenario-to-score pipeline: scenario definition, trace generation, metric computation, and TLRI risk scoring}
    \label{fig:placeholder}
\end{figure}

\subsubsection{Trace structure and notation}
Every trace $i$  is a pair of binary secret labels $s_i$ and timings $t_i$ (in cycles). We represent the data of one scenario as
\[
\mathcal{D}=\{(s_i,t_i)\}_{i=1}^{N}.
\]
The traces are split into two groups:
\[
\mathcal{T}_0=\{t_i:\ s_i=0\},\qquad \mathcal{T}_1=\{t_i:\ s_i=1\},
\]
with $n_0+n_1=N$. We explicitly model timing as a staged process:
\[
t_i \;=\; \underbrace{t_i^{(0)}}_{\text{baseline}}\;\xrightarrow{\text{environment}}\;\underbrace{t_i^{(\mathrm{env})}}_{\text{noise+contention}}\;\xrightarrow{\text{leakage}}\;\underbrace{t_i}_{\text{final observed}}.
\]
This structure is convenient, as it isolates the contribution due to noise and contention as opposed to the contamination by the secret-dependent contribution, the most valuable aspect of leakage analysis.

\vspace{0.25em}
\noindent\textbf{Notation summary (used consistently):}
\begin{itemize}
    \item $B$: baseline cycles for a scheme; $\epsilon$: multiplicative drift; $n$: additive noise; $d$: structured delay.
    \item $\mathcal{T}_0,\mathcal{T}_1$: timing samples conditioned on $s=0$ and $s=1$.
\end{itemize}

\subsection{Secret Sampling and Baseline Timing Model}
\subsubsection{Secret generation model}
Secrets are generated independently using a Bernoulli process:
\[
s_i \sim \mathrm{Bernoulli}(0.5),\qquad i=1,\dots,N.
\]
A balanced parameter of $0.5$ is imposed. This eliminates the effects of class imbalance during hypothesis-testing and simplified the effect sizes into a more understandable form. Independence assumption The trials are independent experiments; secrets do not have a time-dependence. Timing analysis also has binary labeling in common with standard two-class distinguishers.

\subsubsection{Baseline cycles and scheme parameterisation}
For each scheme, a baseline time (cycles) is defined as a constant $B$:
\[
t_i^{(0)} = B,\qquad \forall i.
\]
In this case $B$ means a synthetic average effort of computation and is independent of a given implementation. The environment perturbations (e.g. the scale of Gaussian noise, the means of the exponential delays, the probability of interrupts, the number of blocks to use to aggregate jitter) are also controlled by additional parameters of the scheme. The idea is to have the baseline clear and interpretable in nature, but have the internal environment and leakage dynamics work towards the apparent difference between the state of $\mathcal{T}_0$ and $\mathcal{T}_1$.

\subsection{Environment Noise Models}
\subsubsection{General form: multiplicative drift and additive delays}
Environment effects are modeled using:
\[
t_i^{(\mathrm{env})} = B(1+\epsilon_i) + n_i + d_i.
\]
Multiplicative term $\epsilon_i$ is an approximation of the drift and global frequency variation which is part of DVFS-style drift. The additive noise $n_i$ represents the unstructured noise; and $d_i$ the structured noise (queueing, scheduling, interrupts) that tend to be very heavy tailed. This decomposition is significant since the detectability is a ratio of the separation that is caused by leakage with the cumulative variance and tail behavior of the environment.

\subsubsection{Idle environment (low contention)}
In the idle environment, drift and noise are modeled as:
\[
\epsilon_i \sim \mathcal{N}(0,\sigma_{\mathrm{dvfs}}),
\qquad
n_i \sim \mathcal{N}(0,\sigma_{\mathrm{idle}}),
\qquad d_i=0,
\]
thus,
\[
t_i^{(\mathrm{idle})}=B(1+\epsilon_i)+n_i.
\]
This regime is more like a controlled regime with limited competition giving an indication of relative ease at which to detect small secret-dependent shifts in the event of tight distributions.

\subsubsection{Jitter environment (block-aggregated micro-jitter)}
The jitter environment models accumulation of many small perturbations across conceptual ``basic blocks''. Let $M$ be the number of blocks. For each trace $i$:
\[
x_{i,j}\sim \mathcal{N}\!\left(0,\frac{\sigma_{\mathrm{jitter}}}{\sqrt{M}}\right),\quad j=1,\dots,M,
\qquad
\mathrm{jitter}_i=\sum_{j=1}^{M} x_{i,j},
\]
with increased drift:
\[
\epsilon_i\sim \mathcal{N}(0,1.5\sigma_{\mathrm{dvfs}}),
\]
so that
\[
t_i^{(\mathrm{jitter})}=B(1+\epsilon_i)+\mathrm{jitter}_i.
\]
The  $\sqrt{M}$ scaling maintains the level of total variance with a change in the $M$. Theoretically, $M$ is an approximation of a large number of micro-variations (pipeline stalls, cache effects, predictor noise) adding up in a manner of execution.

\subsubsection{Loaded environment (contention, queueing, interrupts)}
The loaded environment introduces heavy-tailed delays and sporadic interrupts. We sample
\[
q_i \sim \mathrm{Exponential}(\lambda^{-1}),
\qquad
I_i \sim \mathrm{Bernoulli}(p_{\mathrm{int}}),
\qquad
u_i\sim\mathrm{Exponential}(\mu^{-1}),
\qquad
d_i=I_i\cdot u_i,
\]
with stronger noise and drift:
\[
n_i\sim \mathcal{N}(0,2.5\sigma_{\mathrm{idle}}),
\qquad
\epsilon_i\sim \mathcal{N}(0,2\sigma_{\mathrm{dvfs}}).
\]
Thus,
\[
t_i^{(\mathrm{loaded})}=B(1+\epsilon_i)+q_i+d_i+n_i.
\]
The regime is specifically designed to be difficult to distinguish: exponential tail augment overlap and are able to obscure mean shifts, inspiring distributional and dependency measures (KS-$D$, MI) alongside the Welch's test.

\subsection{Secret-Dependent Leakage Injection}
\subsubsection{General leakage operator and strength parameter}
Leakage is introduced after environment noise as an additive secret-dependent term:
\[
t_i = t_i^{(\mathrm{env})} + \Delta(t_i^{(\mathrm{env})}, s_i;\alpha),
\]
where $\Delta(\cdot)$ is determined by the chosen leakage model and and $\alpha>0$ is multiplier of leakage strength. The effect size should grow with the increase in $\alpha$ although it ought to cause the values to decrease the overlap, rise in the KS-$D$ as well as mutual information and climb in the TLRI score. Simulator has optional negative value clipping (as in $t_i\leftarrow \max(t_i,0)$ to assure that having an object leaving the scene at a certain time is physically valid.

\subsubsection{Branch and early-exit comparison leakage (signed shift)}
Branching or early-exit comparisons that depend on secrets (e.g., timing by using a timing-sensitive version of memcmp) are modeled using a symmetric signed shift:
\[
\Delta(t,s;\alpha)=
\begin{cases}
+\alpha\delta & \text{if } s=1,\\
-\alpha\delta & \text{if } s=0,
\end{cases}
\]
where $\delta$ is a constant (cycles). This provides a mean separation that will be expected to be in the order of $2\alpha\delta$ (noise not taken into consideration). It is a model of interpretable (base-level) leakage, whose output is mainly a mean shift that is well explained by Welch in his $t$, and the SNR component within TLRI.

\subsubsection{Division-latency style leakage (binomial event accumulation)}
Division-latency leakage is considered to be modeled as an accumulation of slow events which is secret-dependent. Assume: Let $L$ be opportunities and $\rho(s)$ be the event probability:
\[
E_i \sim \mathrm{Binomial}(L,\rho(s_i)),
\]
with secret-conditioned rates
\[
\rho(1)=\rho_0(1+0.6\alpha),\qquad
\rho(0)=\rho_0(1-0.3\alpha),
\]
and leakage cost
\[
\Delta(t,s;\alpha)=E_i\cdot(\alpha c+\eta_i),
\qquad
\eta_i\sim\mathcal{N}(0,0.25(\alpha c)).
\]
Such a change in distribution form and variance (in addition to the mean) is what causes KS-$D$ and mutual information to be significant metrics, in particular, when things are loaded.

\subsubsection{Cache-index leakage (secret-dependent miss rate)}
Cache-index leakage is modeled via secret-dependent cache miss frequency:
\[
M_i\sim\mathrm{Binomial}(L',\pi(s_i)),
\]
with
\[
\pi(1)=\pi_0+\alpha\Delta_\pi,\qquad
\pi(0)=\max(0,\pi_0-\alpha\Delta_\pi),
\]
and penalty aggregation
\[
\Delta(t,s;\alpha)=M_i\cdot(P+\xi_i),
\qquad
\xi_i\sim\mathcal{N}(0,0.15P).
\]
This directly links leakage to expectation: $\mathbb{E}[M_i|s]=L'\pi(s)$, so the expected timing gap scales with $L'(\pi(1)-\pi(0))P$, whereas the variance scales with the binomial dispersion and penalty noise.

\subsubsection{Large-penalty bit-branch leakage (Frodo-like scaling)}
In large scales of baseline, the simulator uses a large-penalty signed shift:
\[
\Delta(t,s;\alpha)=
\begin{cases}
+\alpha\delta_{\mathrm{big}} & \text{if } s=1,\\
-\alpha\delta_{\mathrm{big}} & \text{if } s=0.
\end{cases}
\]
This is an example of instances of conditional work differentials being associated with large absolute cycle differentials. It still relies on environmental variance and tails and thus the identical suite of metrics are used.

\subsection{Statistical and Information-Theoretic Leakage Metrics}
\subsubsection{Descriptive statistics: mean, variance, and pooled deviation}
We compute per-class sample mean and standard deviation:
\[
\bar{t}_s=\frac{1}{n_s}\sum_{t\in\mathcal{T}_s} t,\qquad
\sigma_s=\sqrt{\frac{1}{n_s-1}\sum_{t\in\mathcal{T}_s}(t-\bar{t}_s)^2},
\qquad s\in\{0,1\},
\]
and pooled deviation
\[
\sigma_p=\sqrt{\frac{\sigma_0^2+\sigma_1^2}{2}}.
\]
These quantities give meaningful information about all the downstream metrics: $\bar{t}_1-\bar{t}_0$ is the observed mean gap, while $\sigma_s$ indicates the noise scale (and typically expands significantly in the loaded regime).

\subsubsection{Welch's t-test (difference in means with unequal variances)}
Welch’s statistic is
\[
t=\frac{\bar{t}_0-\bar{t}_1}{\sqrt{\frac{\sigma_0^2}{n_0}+\frac{\sigma_1^2}{n_1}}}.
\]
The Welch's test works well when leakage causes a shift in mean (e.g., branch leakage), but it is also valid when variances are different. Nonetheless, when the distribution differences lie in the tails or the shape, when the conditions are heavy tailed, Welch $t$ is applied with KS-$D$, effect size, and information measures.

\subsubsection{Kolmogorov--Smirnov statistic (distributional separation)}
The KS statistic is defined as
\[
D=\sup_x \left|F_0(x)-F_1(x)\right|,
\]
where $F_0$ and $F_1$ are empirical CDFs of $\mathcal{T}_0$ and $\mathcal{T}_1$. KS-$D$ captures any separation (mean, variance, skew, tails) and thus is a major key indicator of event-based leakage and loaded environments.

\subsubsection{Cliff's delta (nonparametric effect size)}
Cliff’s delta is
\[
\delta = \Pr(t_0>t_1)-\Pr(t_0<t_1),
\]
with $t_0\sim \mathcal{T}_0$ and $t_1\sim \mathcal{T}_1$. This gives a useful ordering-based effect size in $[-1, 1]$, to the effect size of statistical significance.

\subsubsection{Binned mutual information (dependency in bits)}
Mutual information between secret $S$ and timing $T$ is estimated by binning timing into $B$ bins:
\[
I(S;T)=\sum_{s\in\{0,1\}}\sum_{b=1}^{B} p(s,b)\log_2\frac{p(s,b)}{p(s)p(b)}.
\]
MI reports in bits, and identifies any kind of dependency structure, which, even the mean-only tests fail to reveal the weaknesses about subtle dependencies.

\subsubsection{Distribution overlap (common mass between timing classes)}
Overlap is approximated from histogram masses as
\[
\mathrm{overlap}=\sum_{b=1}^{B} \min\bigl(p_0(b),p_1(b)\bigr),
\]
where $p_s(b)$ is the estimated probability mass for secret class $s$ in bin $b$. An overlap near $1$ provides an indication of near-identical timing distributions; an overlap near $0$ indicates very much separation.

\subsection{Composite Risk Quantification via TLRI-Style Score}
\subsubsection{Signal-to-noise proxy and metric normalisation}
We compute an SNR proxy:
\[
\sigma_p=\sqrt{\frac{\sigma_0^2+\sigma_1^2}{2}},
\qquad
\mathrm{SNR}=\frac{|\bar{t}_0-\bar{t}_1|}{\sigma_p}.
\]
For MI, we apply bounded scaling:
\[
\widehat{\mathrm{MI}}=\min\left(1,\frac{I(S;T)}{0.5}\right).
\]
Any such normalisations are metric contributions that are fused to avoid any single metric dominance because of scale.

\subsubsection{Raw aggregation and logistic mapping}
We combine evidence into a raw score:
\[
\mathrm{raw}=
0.9\cdot \mathrm{SNR}
+1.3\cdot D
+1.1\cdot |\delta|
+1.2\cdot (1-\mathrm{overlap})
+0.9\cdot \widehat{\mathrm{MI}},
\]
and map to $[0,1]$ using:
\[
\mathrm{TLRI}=\frac{1}{1+\exp\left(-(\mathrm{raw}-1.5)\right)}.
\]
The single bounded indicator by TLRI is applicable to rank scenarios but is nonetheless based on more than one independent leakage view.

\subsection{Sample-Size Sensitivity and Stability Analysis}
\subsubsection{Motivation: sample complexity of leakage detection}
Leakage evidence will be affected by the number of traces. Certain situations are well separated with a small $N$ whereas others need very large samples due to noise, tails or other lurking dependencies. We measure this by computing metric sweeps at growing prefix size to yield stability curves as in the form of the $\mathrm{TLRI}(N)$ and $D(N)$.

\subsubsection{Sweep construction and metric recomputation}
For sweep points $N_k$ between a minimum threshold and $N_{\max}$, we define:
\[
\mathcal{D}_{N_k}=\{(s_i,t_i)\}_{i=1}^{N_k},
\qquad
\mathrm{Metrics}(N_k)=\Phi(\mathcal{D}_{N_k}),
\]
where $\Phi(\cdot)$ is the full metric pipeline (Welch, KS, $\delta$, MI, overlap, TLRI). Random permutation of the traces with the fixed seed is done to eliminate ordering artefacts, after which measures are re-computed on prefixes to mirror the accumulation of evidence with sample size.

\subsection{Reproducibility Controls and Scope}
\subsubsection{Random seed control and deterministic regeneration}
All the stochastic variables (secret sampling, Gaussian noise, exponential random delay, binomial random number of events) have a deterministic random seed. Thus, a certain configuration will reproducibly recreate the identical traces and identical metrics. Before the main simulation, a warming up using the PRNG state might be performed.

\subsubsection{Scope, assumptions, and interpretive limits}
This synthetic and parameterised simulator is very specific. This means that reported values are not of a particular codebase or hardware platform, but are distinguishable as modeled (environment + leakage). This methodology will be used in professional risk modelling: It offers a systematic structure into which the behaviour of various leakage mechanisms under different execution regimes can be compared, and on which cases particular situations are prioritised with respect to further empirical validation.

\section{Results}
The results of our synthetic timing side-channel risk simulator are reported here on three families of post-quantum KEMs (Kyber, Saber, Frodo), three execution platforms (idle, jitter, loaded), and three leakage models. In each case, timing traces of two equally likely secret classes were generated and compared with Welch $t$-test, KS distance, Cliff, and the mutual information, distribution overlap, and the composite TLRI score.

\subsection{Overall leakage risk across scenarios}

\begin{figure}[H]
    \centering
    \includegraphics[width=\linewidth]{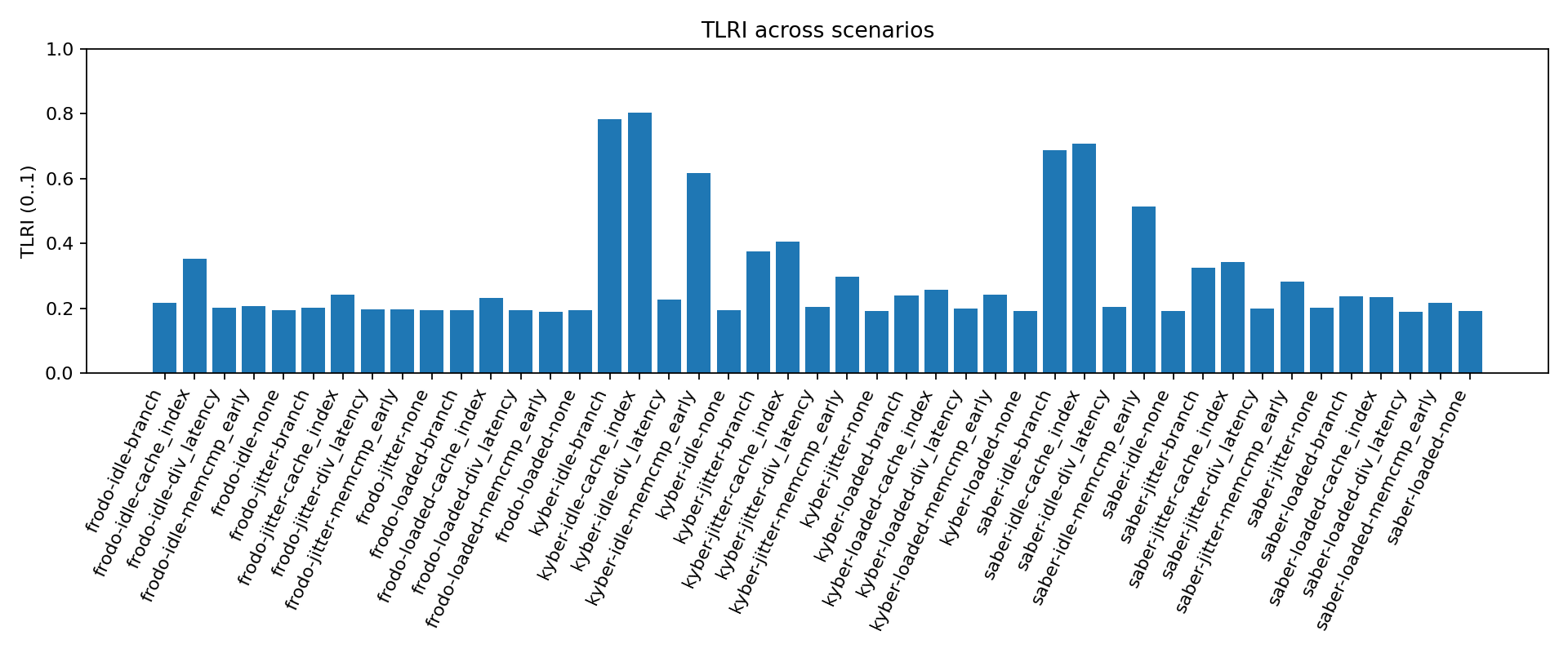}
    \caption{TLRI scores across all simulated scenarios (scheme $\times$ environment $\times$ leakage type). Higher TLRI indicates stronger timing distinguishability between secret classes.}
    \label{fig:tlri_overview}
\end{figure}

Figure~\ref{fig:tlri_overview} summarizes TLRI values across all simulated scenarios. Two dominant trends are immediately visible:

\begin{itemize}
    \item \textbf{Environment strongly affects detectability:} the idle environment consistently produces higher TLRI values than jitter or loaded settings.
    \item \textbf{Leak type dominates risk:} cache-index leakage and branch-based leakage yield the strongest distinguishability in multiple schemes, while div-latency shows consistently weaker effects under the chosen parameters.
\end{itemize}

\subsection{Baseline vs worst-case leakage comparison}

To contextualize the leakage strength, Table~\ref{tab:key_summary} compares (i) the baseline no-leak case with (ii) the strongest leakage case per scheme/environment. In all schemes, baseline TLRI is close to a low-risk bottom, whereas in lazy contexts, worst-case leakage is much more detectable.

\begin{itemize}
    \item \textbf{Kyber exhibits the highest peak risk:} its idle cache-index scenario reaches TLRI $\approx 0.80$.
    \item \textbf{Saber shows similar behavior:} its strongest idle cache-index scenario reaches TLRI $\approx 0.71$.
    \item \textbf{Frodo remains lower overall:} even its worst-case scenario remains below $\approx 0.36$, consistent with higher intrinsic noise and large baseline timing magnitude.
\end{itemize}

\begin{table}[H]
\centering
\caption{Baseline (no leak) vs. worst-case TLRI per (scheme, environment).}
\label{tab:key_summary}
\begin{tabular}{lllrrr}
\hline
Scheme & Env. & Base TLRI & Worst leak & Worst TLRI & $\Delta$TLRI \\
\hline
frodo & idle   & 0.195 & cache\_index  & 0.353 & 0.158 \\
frodo & jitter & 0.192 & cache\_index  & 0.243 & 0.051 \\
frodo & loaded & 0.190 & cache\_index  & 0.230 & 0.040 \\
kyber & idle   & 0.195 & cache\_index  & 0.803 & 0.608 \\
kyber & jitter & 0.190 & cache\_index  & 0.405 & 0.215 \\
kyber & loaded & 0.195 & cache\_index  & 0.259 & 0.064 \\
saber & idle   & 0.191 & cache\_index  & 0.706 & 0.515 \\
saber & jitter & 0.193 & cache\_index  & 0.339 & 0.146 \\
saber & loaded & 0.191 & cache\_index  & 0.238 & 0.047 \\
\hline
\end{tabular}
\end{table}

\subsection{Top leakage scenarios by TLRI}

Table~\ref{tab:top_scenarios} shows the most robust leakage scenarios in the world according to TLRI. Cache-index leakage under idle environments and branch-type leakage follows closely as the top in the ranking of cases. These measurements show that a regular secret-dependent timing-modulation will yield statistically separable timing- distributions in moderate signal-noise levels.

\begin{table}[H]
\centering
\caption{Top leakage scenarios ranked by TLRI (excluding no-leak baseline).}
\label{tab:top_scenarios}
\begin{tabular}{lllrrrr}
\hline
Scheme & Env. & Leak & TLRI & KS $D$ & $|\delta|$ & MI (bits) \\
\hline
kyber & idle & cache\_index & 0.803 & 0.407 & 0.548 & 0.1816 \\
kyber & idle & branch & 0.784 & 0.397 & 0.529 & 0.1678 \\
saber & idle & cache\_index & 0.706 & 0.339 & 0.465 & 0.1296 \\
saber & idle & branch & 0.687 & 0.329 & 0.450 & 0.1193 \\
kyber & idle & memcmp\_early & 0.618 & 0.287 & 0.396 & 0.0927 \\
kyber & idle & div\_latency & 0.404 & 0.178 & 0.254 & 0.0330 \\
kyber & jitter & cache\_index & 0.405 & 0.194 & 0.278 & 0.0392 \\
kyber & jitter & branch & 0.377 & 0.178 & 0.254 & 0.0337 \\
saber & jitter & cache\_index & 0.339 & 0.160 & 0.229 & 0.0257 \\
saber & jitter & branch & 0.322 & 0.150 & 0.215 & 0.0237 \\
frodo & idle & cache\_index & 0.353 & 0.170 & 0.243 & 0.0308 \\
frodo & idle & branch & 0.347 & 0.167 & 0.239 & 0.0298 \\
frodo & jitter & cache\_index & 0.243 & 0.100 & 0.144 & 0.0106 \\
kyber & loaded & cache\_index & 0.259 & 0.108 & 0.155 & 0.0118 \\
saber & loaded & cache\_index & 0.238 & 0.095 & 0.136 & 0.0092 \\
\hline
\end{tabular}
\end{table}

\subsection{Distribution-level evidence (per scheme)}

We present only a single representative worst-case leakage scenario in each scheme as a histogram+KDE overlap, and ECDF separation, and violin curves in order to avoid congesting the mathematical expression with numbers. Such plots give an intuitive justification of the statistical measures.

\subsubsection{Kyber: strongest distinguishability under cache-index leakage}

With the idle cache-index leakage model (TLRI $\approx 0.80$), Kyber demonstrates the best observed TLRI. In both cases, the reduction of the density overlap and divergence in ECDF is seen, which is expected because the KS distance is moderate and the effect size is large.

\begin{figure}[H]
    \centering
    \includegraphics[width=0.32\linewidth]{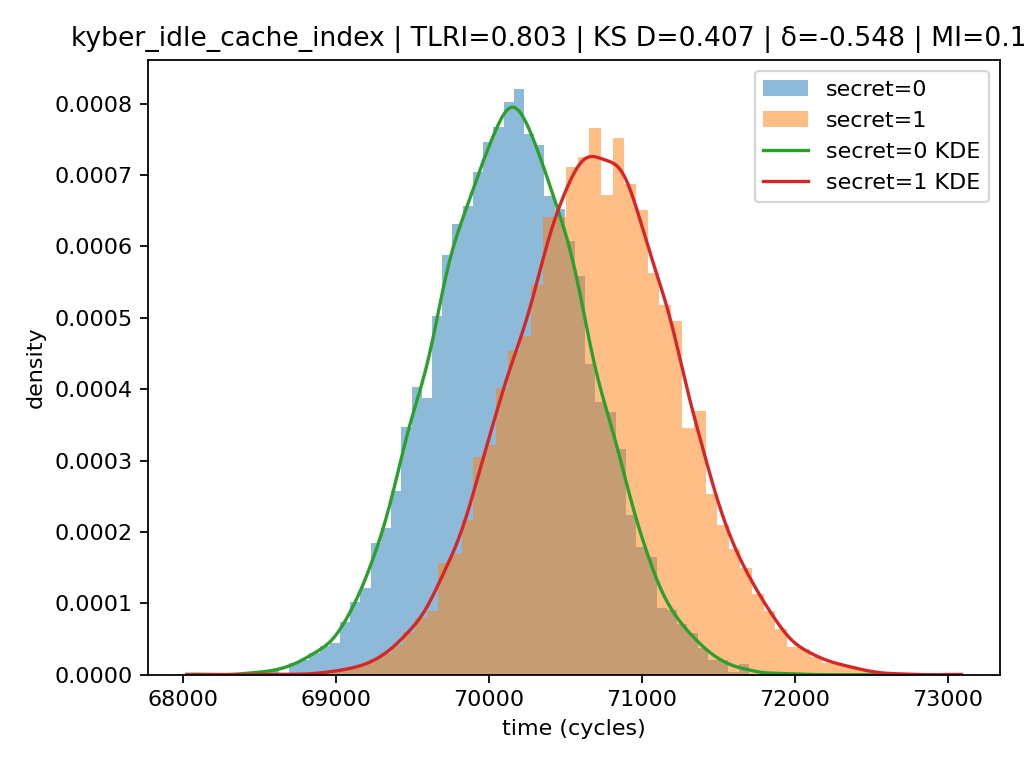}
    \includegraphics[width=0.32\linewidth]{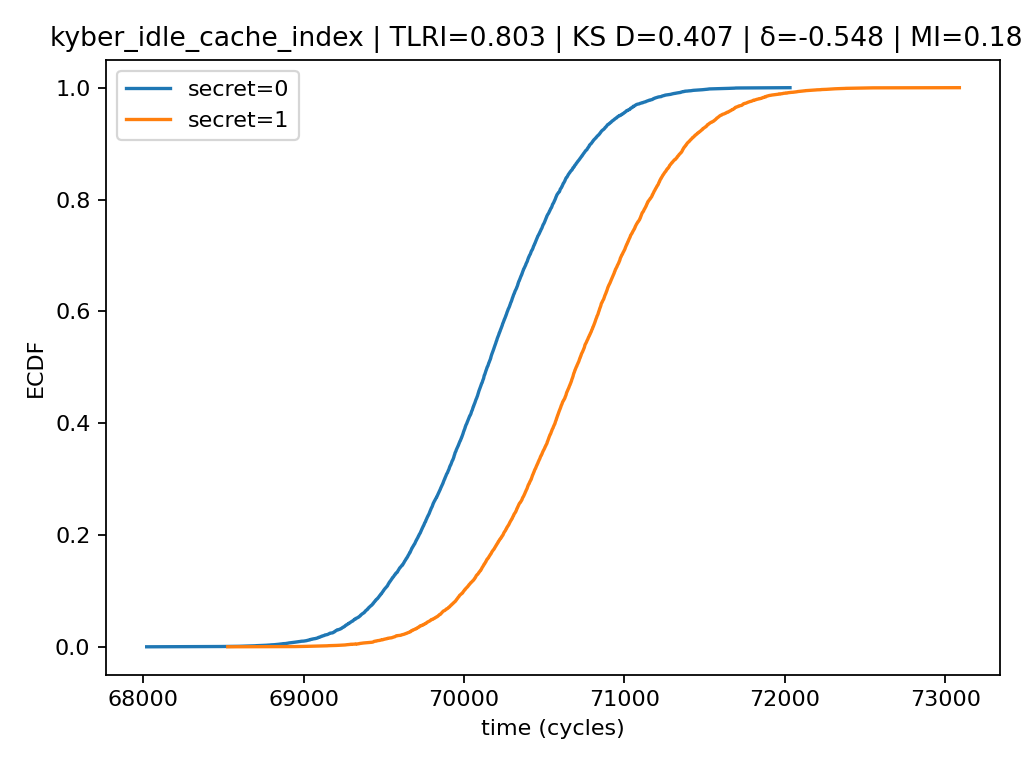}
    \includegraphics[width=0.32\linewidth]{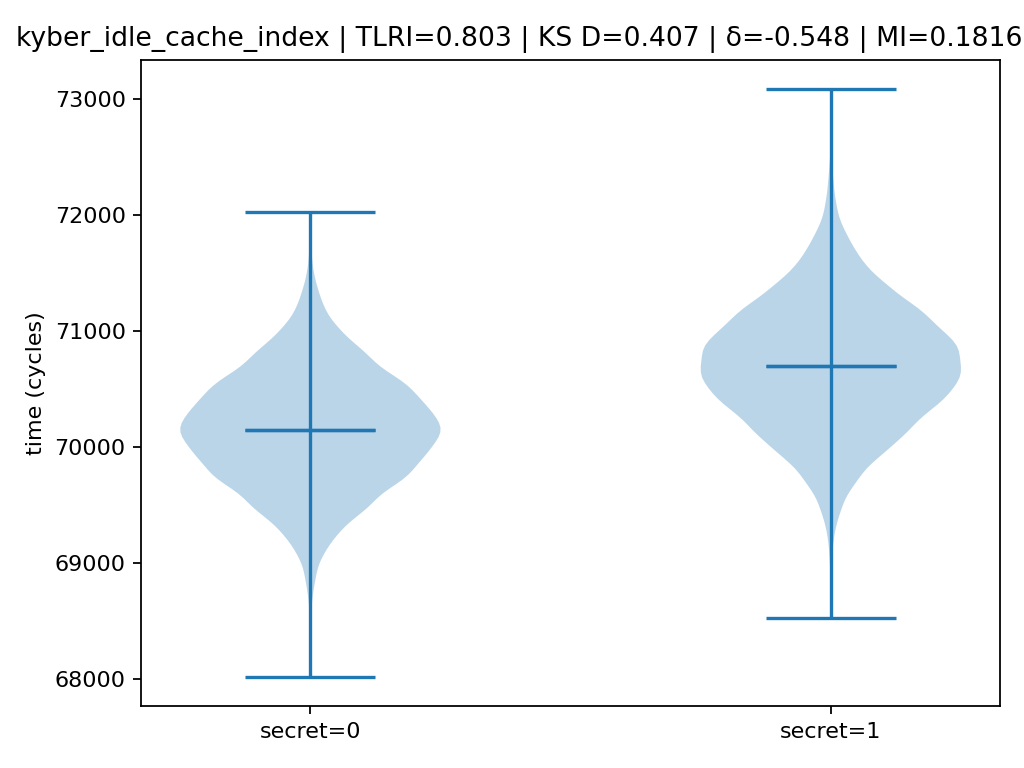}
    \caption{Kyber worst-case scenario (idle + cache-index leakage). Left: histogram+KDE. Middle: ECDF. Right: violin plot.}
    \label{fig:kyber_case}
\end{figure}

\subsubsection{Saber: comparable leakage behavior with slightly reduced magnitude}

The best scenario of Saber is also the one with idle cache-index leakage (TLRI $\approx 0.71$), as well. Although it remains quite distinguishable, Saber is a little more overlapped and lower in the degree of effects as compared to Kyber, resulting in a lower composite risk score.

\begin{figure}[H]
    \centering
    \includegraphics[width=0.32\linewidth]{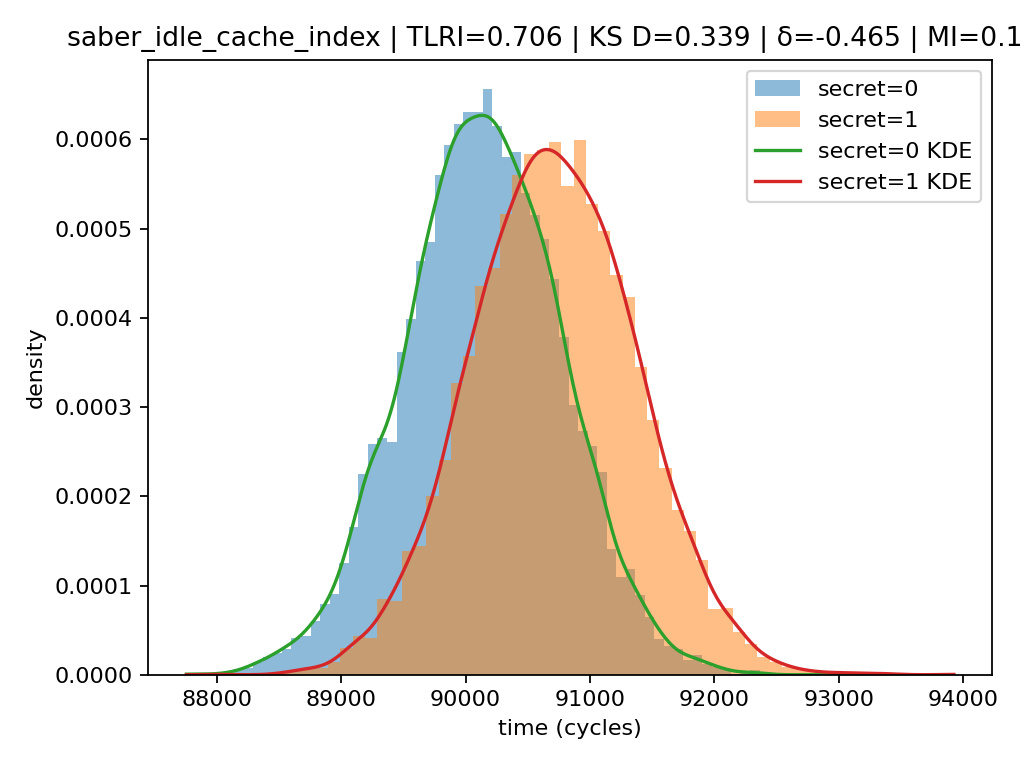}
    \includegraphics[width=0.32\linewidth]{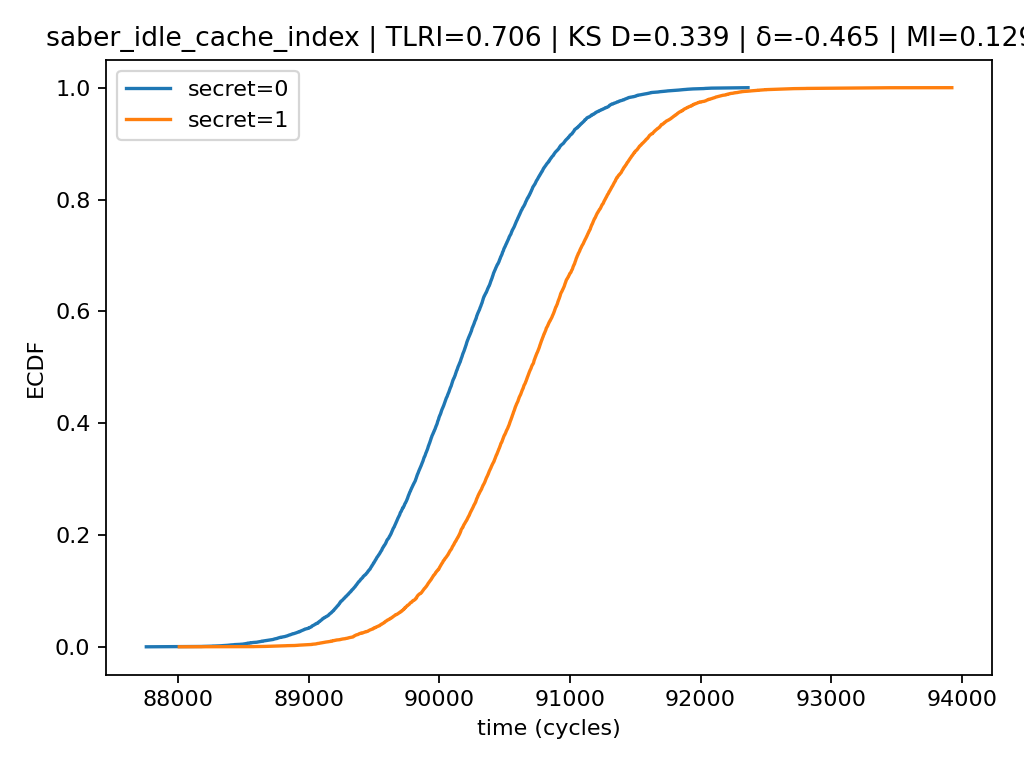}
    \includegraphics[width=0.32\linewidth]{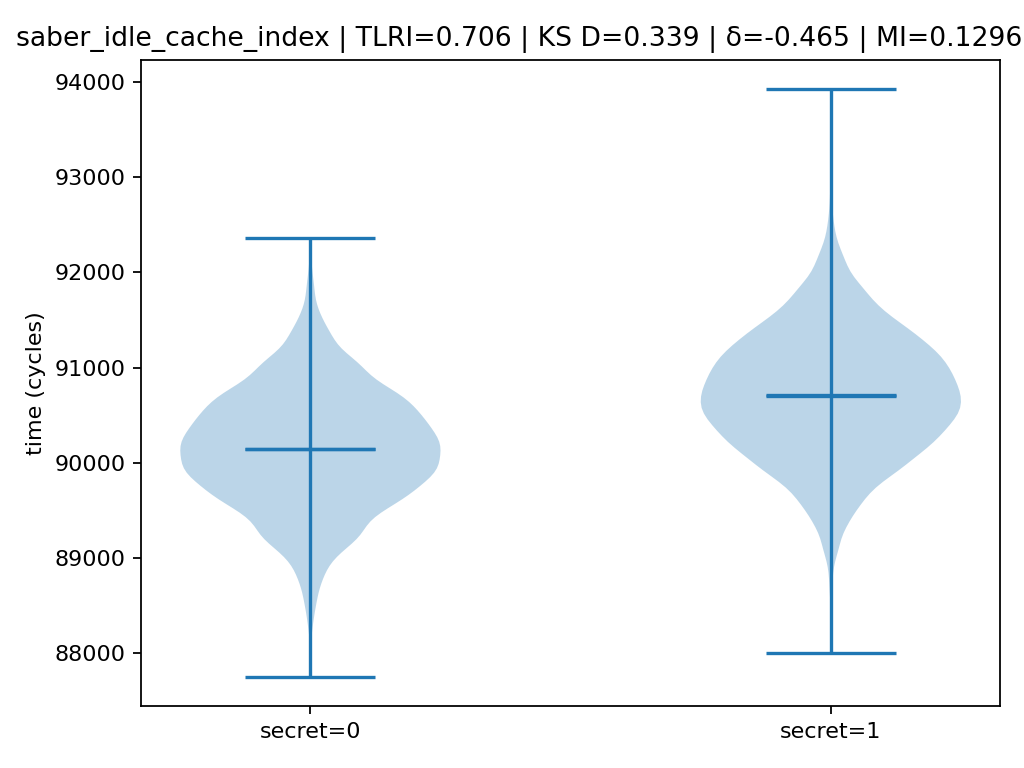}
    \caption{Saber worst-case scenario (idle + cache-index leakage). Left: histogram+KDE. Middle: ECDF. Right: violin plot.}
    \label{fig:saber_case}
\end{figure}

\subsubsection{Frodo: reduced TLRI due to high noise scale and large baseline}

Frodo has the smallest maximum TLRI of all the simulated schemes (peak TLRI below $\approx 0.36$). Although the leakage is there, the distributions are still more similar with each other, in accordance with its large timing variance and heavy-tailed character.

\begin{figure}[H]
    \centering
    \includegraphics[width=0.32\linewidth]{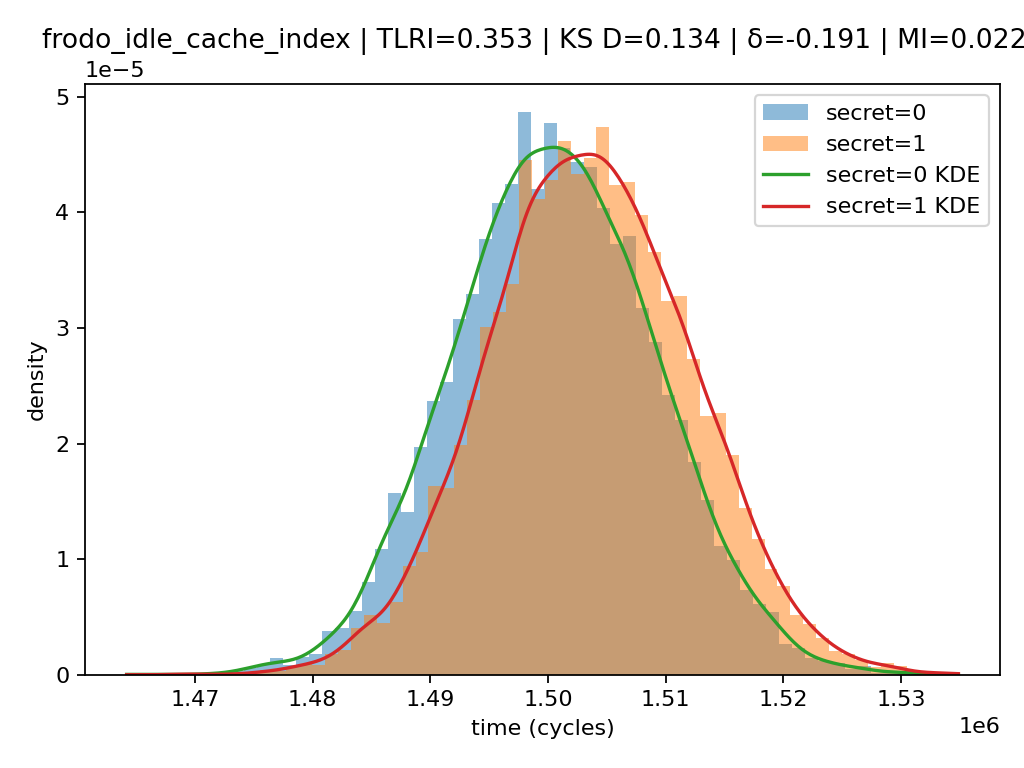}
    \includegraphics[width=0.32\linewidth]{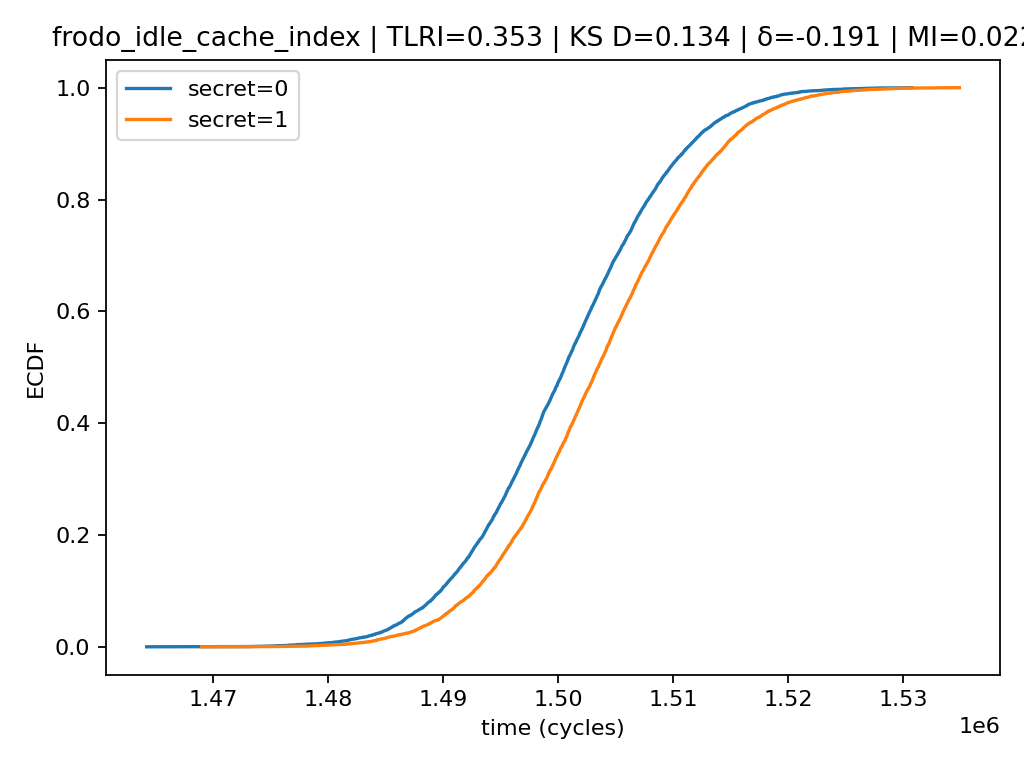}
    \includegraphics[width=0.32\linewidth]{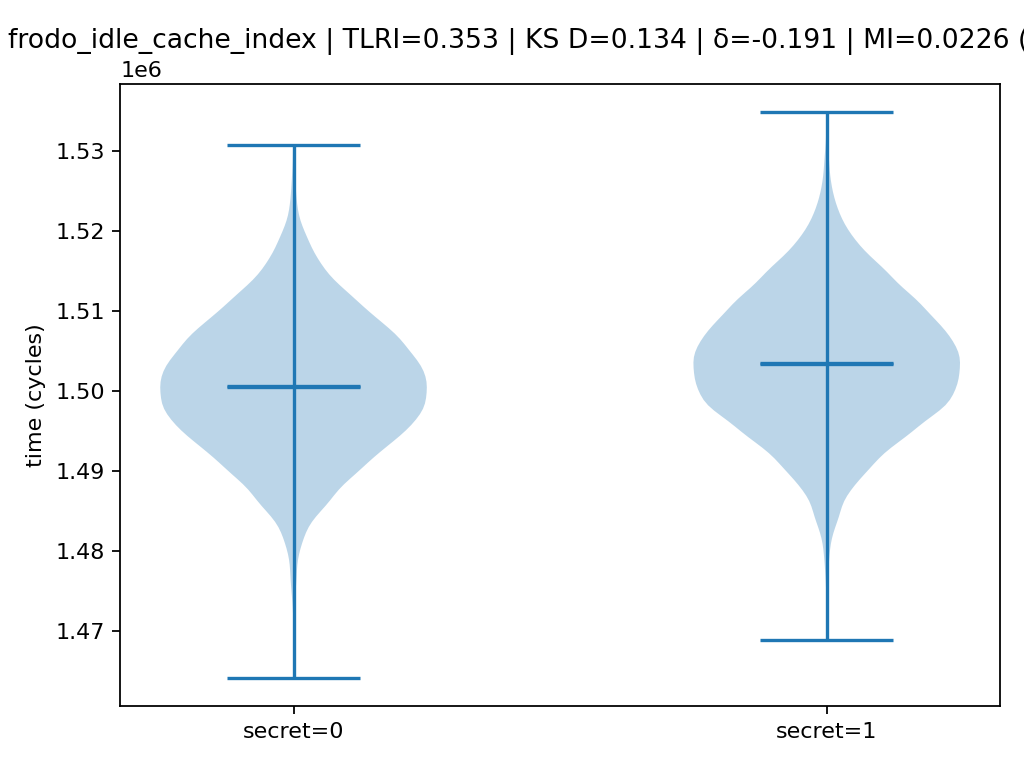}
    \caption{Frodo worst-case scenario (idle + cache-index leakage). Left: histogram+KDE. Middle: ECDF. Right: violin plot.}
    \label{fig:frodo_case}
\end{figure}

\subsection{Key empirical observations}

Overall, the results support the following conclusions:

\begin{itemize}
    \item \textbf{Idle conditions maximize detectable leakage}, while loaded and jitter environments reduce distinguishability by increasing variance and overlap.
    \item \textbf{Cache-index leakage is consistently the strongest leak model} under the simulated parameters, producing the highest TLRI values across schemes.
    \item \textbf{Fast schemes (Kyber, Saber) show higher risk scores than Frodo} under equivalent leakage assumptions, because relative timing separation is larger compared to their noise scale.
\end{itemize}

\section{Conclusion}
Timing side-channels are also a practical attack on cryptographic tools, and with post-quantum cryptography, this is more of a concern since lattice based constructions may impose variability based on secrets in timing through complex arithmetic and control structures. Given that the timing measurements under real conditions are also affected by the noise of the environment (i.e.: scheduling effects, contention, heavy-tailed delays) this paper hypothesizes a scenario-driven statistical risk model that considers timing leakage a distributional distinguishability challenge in controlled execution conditions. Our synthetic traces of two secret classes among idle, jitter, and loaded settings and an array of leakage behaviors are quantified by Welch t-test, KS distance, Cliff delta, mutual information, and distribution overlap and combined to yield a TLRI-style composite score to rank among homogenous scenarios in the same manner. In representative lattice-based KEM families (Kyber, Saber, Frodo) idle conditions are typically best separable, and jitter and loaded regimes maximize the risk by maximizing variance and overlap; cache-index, and branch-style leakage are often the most effective indicators of risk, and more risky schemes can be tried on faster platforms with equal leakage assumptions.

\section*{Conflict of Interest}
The authors declare no conflict of interest.

\end{document}